# Deep-UV Silicon Polaritonic Metasurfaces for Enhancing Biomolecule Autofluorescence and Two-Dimensional Material Double-Resonance Raman Scattering

*Bo-Ray Lee, Mao Feng Chiang, Pei Ying Ho, Kuan-Heng Chen, Jia-Hua Lee, Po Hsiang Hsu, Yu Chieh Peng, Jun-Yi Hou, Shih-Chieh Chen, Qian-Yo Lee, Chun-Hao Chang, Bor-Ran Li, Tzu-En Lin, Chieh-Ting Lin, Min-Hsiung Shih, Der-Hsien Lien, Yu-Chuan Lin, Ray-Hua Horng, Yuri Kivshar, Ming Lun Tseng\**

B.-R. Lee, M. F. Chiang, P. Y. Ho, K.-H. Chen, J.-H. Lee, P. H. Hsu, Y. C. Peng, J.-Y. Hou, S.-C. Chen, D.-H. Lien, R.-H. Horng, M. L. Tseng*
Institute of Electronics
National Yang Ming Chiao Tung University
Hsinchu 300093, Taiwan
E-mail: mltseng@nycu.edu.tw

Q. -Y. Lee, Y. -C. Lin
Department of Materials Science and Engineering
National Yang Ming Chiao Tung University
Hsinchu 300093, Taiwan

C. -H. Chang, B. -R. Li
Institute of Biomedical Engineering
National Yang Ming Chiao Tung University
Hsinchu 30093, Taiwan

T. -E. Lin
Graduate Institute of Applied Mechanics
National Taiwan University
Taipei 106, Taiwan

C. -T. Lin
Department of Chemical Engineering
National Chung Hsing University
Taichung 402, Taiwan

M. -H. Shih
Research Center for Applied Sciences
Academia Sinica
Taipei 11529, Taiwan

Y. Kivshar
Nonlinear Physics Centre
Research School of Physics
Australian National University
Canberra ACT 2601, Australia

**Abstract**

High-performance deep ultraviolet (DUV) spectroscopy is crucial in driving innovations for biomedical research, clinical diagnosis, and material science. DUV resonant nanostructures have shown capabilities for significantly improving spectroscopy sensitivity. However, they encounter significant challenges in practical applications, including instability due to oxidation and light-induced damage, and the strong photoluminescent noise background from their constituent materials. We propose an efficient and robust DUV spectroscopy platform based on the polaritonic properties in all-dielectric silicon (Si) metasurfaces. Unlike conventional dielectric metasurfaces that rely on Mie-type modes, our approach leverages the polaritonic resonances in Si nanostructures—a striking yet underexplored property driven by interband transitions in the DUV regime—for nanophotonic sensing. A polaritonic Kerker-type void metasurface providing strong near-field enhancement localized on the surface was designed and fabricated. The metasurface facilitates double-resonance Raman scattering, a process that reveals key information about lattice dynamics and electronic structures, for analyzing 2D semiconductor monolayers. It also demonstrates superior stability in solvents and enhances biomolecule autofluorescence. These capabilities demonstrate the versatile potential of Si metasurfaces as a scalable, robust platform for interdisciplinary DUV spectroscopy applications, including advanced biomedical research and the investigation of emerging nanomaterials.

1. Introduction

Deep ultraviolet (DUV) light, with wavelengths from 200 to 300 nm, strongly interacts with biomolecules and materials, creating distinct spectral features in Raman scattering[1], autofluorescence[2-3], and absorption[4] measurements that cannot be observed with visible light counterparts. These interactions offer a valuable platform for advancing biomedical analysis[2-3, 5], clinical screening[6], and nanomaterial investigation[1]. In recent years, significant efforts have been made to employ nanophotonic structures, such as metasurfaces, to advance molecular spectroscopy[7-12]. These nanostructures enhance light-matter interactions by confining light to the nanoscale, amplifying weak optical signals, and improving measurement sensitivity.

In the DUV regime, only a limited number of materials are suitable for metasurfaces. Aluminum (Al) is widely used in sensing plasmonic nanostructures[13-14]. Despite recent advancements[15-17], issues such as instability and fragility remain problematic. Continuous oxidation[18] and light-induced damage[19-21] affect Al nanostructures in various sensing applications, including in-flow molecular sensing[19] and Raman spectroscopy[22]. While protective coatings mitigate these issues, they also reduce near-field enhancement and increase fabrication costs due to complex deposition processes. Alternatives like rhodium[23] are hindered by limited availability and fabrication challenges. Another approach involves using Mie-type metasurfaces made of dielectric materials like AlN[8], $HfO_2$[24], and diamond[25]. However, these materials often exhibit strong photoluminescence under DUV illumination[26-28], which generates significant background noise during spectroscopy measurements. In many cases, the photoluminescence signals from metasurfaces made of these materials can be far stronger than those from the analytes, greatly reducing their effectiveness in sensing applications. Although there are methods for fabricating diamond nanostructures, they are not yet fully developed. To advance DUV metasurface-based sensing technologies, a more robust, stable, and scalable platform is essential.

  We propose that Si can be the striking solution for DUV nanophotonic sensing applications. Si is commonly regarded as a dielectric and widely used in optical communication, photovoltaics, and nanophotonics in the visible and IR regions[29-35]. However, it exhibits a significant shift in properties in the DUV region. Strong interband transitions[36] in Si enable the excitation of surface polaritons—confined electromagnetic modes[37]—in its nanostructures. Unlike conventional dielectric metasurfaces that rely on Mie-type modes with fields usually concentrated inside the resonators[8], the polaritonic resonances in Si metasurfaces effectively confine the resonant field on the device surface at the subwavelength scale. The exposed and enhanced field is highly beneficial for achieving high-performance DUV molecular spectroscopy. Additionally, without specific treatments, Si does not produce pronounced photoluminescence in the UV and visible bands, making it an ideal low-background-noise material platform for sensing spectroscopy applications. Although several works have investigated the resonance properties of Si nanostructures[38-41], the potential for the applications of the relevant DUV surface-enhanced spectroscopy remains largely unexplored. By combining the ability to generate strong resonant surface field with excellent stability, neglectable photoluminescence background, and well-established fabrication processes, Si offers a striking platform for DUV sensing and spectroscopy.

  Here, we explore the potential of unconventional resonances in Si for a range of DUV metasurface-enhanced molecular spectroscopy applications. To enhance light-matter interactions between DUV light and analytes, we designed and fabricated a Kerker-resonant [42-43] void metasurface, capable of generating strong field enhancement at ~266 nm. As a proof of concept, the Si metasurface successfully enhances the photoluminescence of biomolecules over the entire emission spectrum. The metasurface also facilitates the resonance Raman scattering of monolayer two-dimensional (2D) semiconductors. In addition, the metasurface demonstrates superior stability in commonly used biochemical solvents, manifesting its feasibility for biomolecular spectroscopy applications in such environments. The results confirm the versatility of the reported Si metasurface for DUV sensing and analysis applications in biomedical research and material science.

## 2. Property change of Si in the DUV

Surface polariton resonances are surface-bound electromagnetic modes capable of providing significant resonant fields. Previously, they are observed in material systems such as metals[44], polar crystals[45], and 2D materials[46]. A lesser-known but exciting fact is that Si, one of the most earth-abundant elements, can also exhibit strong polaritonic resonances. The change in the optical properties of Si can be clearly verified from its wavelength-dependent optical constants. **Figure 1**a presents the real (Re($\varepsilon$), purple curve) and imaginary (Im($\varepsilon$), blue curve) components of Si's permittivity from 200 nm to 500 nm, derived from experimental data in Ref.[47]. The real part remains positive in the near-UV (300 to 400 nm) and visible ranges, confirming its dielectric nature. However, below 400 nm, the imaginary part, which reflects the material absorption, rises sharply due to Si's interband transitions. Since the complex dielectric function must follow the Kramers-Kronig relations[48], the interband transitions creating an absorption band in the imaginary components lead to a corresponding decline in the real components. With strong oscillation strength of the transitions, the real part turns negative at photon energies around the peak of the imaginary part. Negative permittivity in Si is evident below 295 nm, suggesting that localized surface polaritons—requiring opposite permittivity signs at the interface[37] — can be excited in Si nanostructures under DUV illumination (see Supporting Information for a more detailed discussion). To further verify, we investigated the optical properties of a metasurface consisting of a Si nanorod array, where the design is shown in Figure 1b, by using FDTD-based commercial software Lumerical. As can be identified in the electric field distribution (Figures 1c-d) and the spectra (Figure 1e), the Si metasurface shows two resonance modes—one in the DUV and the other in the visible, respectively. As shown in Figure 1c, strong field enhancement and confinement occur at the edge of the Si nanorod, indicating a surface-bound dipolar resonance at $\lambda = 273$ nm. The surface localization of the resonance can be further verified by checking the cross-sectional electric field profile along the long axis of the Si nanorod (the bottom panel of Figure 1c). It is found that the attenuation depth for the field in the free space and in the Si nanorod are ~4.5 nm and ~2.2 nm, respectively. This light confinement at the deep subwavelength scale (~0.016 $\lambda$ at the Si/air interface) is a typical characteristic of the polaritonic resonance. On the contrary, in the visible range, at 408 nm (Figure 1d), the mode shows a typical characteristic of Mie-type resonance that field anti-nodes (as shown in the bottom panel of Figure 1d) form in the nanorod, similar to the observation in the previous works of Mie-type metasurfaces[8, 29]. The comparison and discussion of the results in Figure 1 clearly show the property transition of Si in the DUV regime and the impact on the resonance properties of the device.

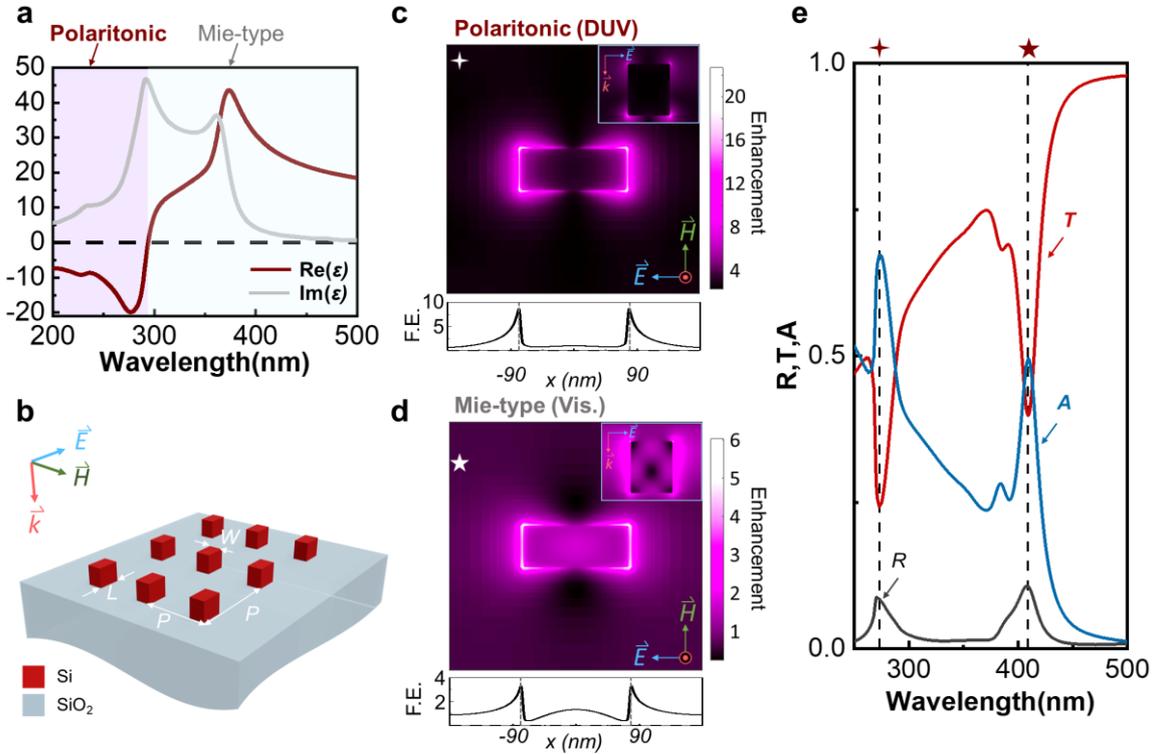

**Figure 1.** Property transition of Si. a) The complex dielectric constants of Si derived from data in Ref. [47]. b) Geometric parameters of the Si nanorod metasurface. *P*: 180 nm, *L*: 100 nm, *W*: 30 nm, *H*: 80 nm. $\vec{E}$: electric field; $\vec{H}$: magnetic field; $\vec{k}$: direction of the incident light. c) The field distribution of the Si nanorod at 273 nm. Bottom panel: The electric field enhancement along the long axis of the rod. It is plotted at the middle of the rod's short axis and at a height of 1 nm from the rod's bottom. The dashed lines indicate that boundaries between the rod and the background. F.E.: field enhancement. The corresponding wavelength is marked with a cross (✦) in (e). d) The field distribution of the Si nanorod at 405 nm. Bottom panel: The electric field enhancement along the long axis of the rod. Bottom panel: The electric field enhancement

along the long axis of the rod. It is plotted at the middle of the rod's short axis and at a height of 1 nm from the rod's bottom. The dashed lines indicate that boundaries between the rod and the background. F.E.: field enhancement. The corresponding wavelength is marked with a star (★) in (e). e) Simulation spectra of the metasurface. *R*: reflectance, *T*: transmittance, *A*: absorbance.

## 3. Sensing chip design and nanofabrication

We further designed a Si metasurface capable of generating strong field enhancement for DUV molecular sensing. **Figure 2**a shows a schematic of the metasurface. It consists of an array of fish-shaped voids, with the subwavelength pattern etched into a Si substrate to a depth (*D*) of 50 nm (Figure 2b). The unit cell design can be considered as a result of overlapping a circular void and a fan-shaped void, and the polaritonic enhancement provided by the metasurface can be effectively tuned by the geometric parameters of the void. Figure 2c illustrates the dependence of the field enhancement (upper panel) and reflectance (bottom panel) on the expanding angle of the unit cell's tail (denoted as *θ* in Figure 2b). We aimed to achieve a design showing a resonance at ~266 nm to match the wavelength of the excitation laser used for the surface-enhanced spectroscopy applications in this paper. A resonant dip in the metasurface is identified in the simulation, and the field enhancement progressively increases as the expanding angle decreases. For sample fabrication consideration, we choose *θ* = 90°. This angle avoids small interspaces between the unit cells that may cause fabrication issues, while maintaining sufficient field enhancement in the metasurface. Consequently, the proposed metasurface provides strong DUV field enhancement without requiring ultrasmall feature sizes (*e.g.*, few nanometer gaps between unit cells) that are usually needed for gap plasmonic devices[49-51]. In our design, the minimal feature size of the metasurface is approximately 85 nm, making it feasible to use standard fabrication processes such as electron beam (e-beam) lithography, nanoimprint lithography, and DUV lithography. The resulting metasurface shows a resonant dip in the reflectance spectrum at ~266 nm. The near-field intensity enhancement, defined as $|E/E_0|^2$, where $|E_0|$ is the incident field amplitude, exceeds 700-fold and is majorly located on the surface of the unit cell's neck at this wavelength (inset of the top panel in Figure 2c). This value is comparable to several previously reported DUV Al plasmonic sensors[16, 52].

To gain further insight into the resonance properties of the metasurface, we calculated multipole decomposition[53-55]. The analysis was performed by simulating the electric displacement current density within the metasurface and applying the formulas described in Ref. [53] (see Supporting Information). Figure 2(b) presents the scattering powers for the electric dipole (ED), magnetic dipole (MD), electric quadrupole (EQ), and magnetic quadrupole (MQ). Indeed, in accordance with Babinet's principle, nanostructures that perforated into films can simultaneously show strong ED and MD resonances[56-57]. At ~266 nm, the ED and MD resonances are of equal strength, providing insight into the metasurface's strong light confinement and field enhancement. Since ED and MD are always out of phase for backward scattering[43, 58], their equal strength will result in the significant suppression of the device's reflection (see Supporting Information for a more detailed discussion). This property satisfies the first generalized Kerker condition[42-43], this concept is also used for the realization of the Huygens' resonance for transmissive metasurfaces[59-61]. The combination of the opaque property of the void-type design and this Kerker-type resonance greatly enhances the light confinement and field strength of the metasurface. In addition, we verify that the working wavelength of the Si metasurface's polaritonic wavelength can be effectively tuned from 284 nm to 184 nm by changing the unit cell's geometric parameters, see the result in Supporting Information. We fabricated the sample on a Si chip using e-beam lithography followed by a dry etching process. For spectral characterization, we built a DUV-Visible micro-spectrometer. Figure 2e presents the reflectance spectrum of fabricated Si metasurface, where the inset shows the scanning electron microscopic (SEM) image of the sample. In the measurement, similar to the simulation condition, the polarization is aligned along the neck of the fish-like voids to ensure the excitation of the resonance. This sample shows a resonance dip, similar to the observation in the simulation results, at ~266 nm, verifying its resonance characteristic at the target wavelength. Further details, including the fabrication process, setup of the micro-spectrometer, effects of unit cell numbers and symmetry, the metasurface's property for different polarization states, a polarization-insensitive design, and the correction for proximity effect for the e-beam exposure, are discussed in the **Experimental Section** and Supporting Information. Overall, the highly confined and enhanced surface field on the surface will be very helpful in boosting up the light-matter interactions between the DUV light and the analytes, resulting in amplified spectroscopy signals in the observation.

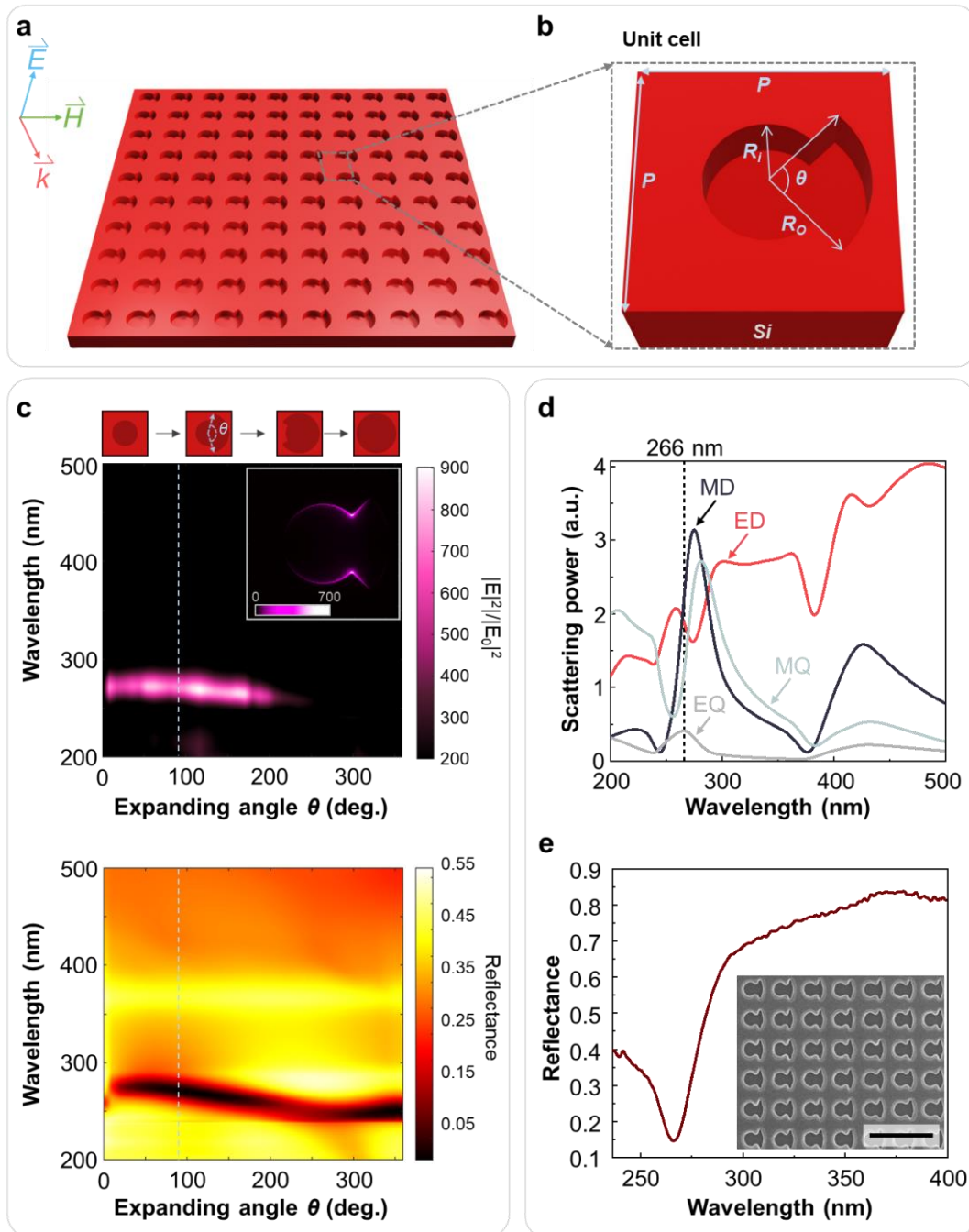

**Figure 2.** Design and analysis of surface-enhanced metasurface. (a)-(b) Metasurface design. *P*: 240 nm, $R_i$: 60 nm, $R_O$: 100 nm, *θ*: 90° The depth of the void: 50 nm. $\vec{E}$: electric field; $\vec{H}$: magnetic field; $\vec{k}$: direction of the incident light. c) Dependence of the electric field enhancement (top panel) and reflectance (bottom panel) on the expanding angle of the unit cell's tail. Dash line indicates the position of 90°. Inset: Simulated electric near-field intensity enhancement $|E/E_0|^2$. $E_0$: the electric field of the incident light. d) Multipole decomposition calculation. ED: electric dipole, MD: magnetic dipole, EQ: electric quadrupole, MQ: magnetic

quadrupole. Dash line indicates the position of 266 nm. e) Experimental reflectance spectrum of the Si metasurface. Inset: SEM image of the Si metasurface. Scale bar: 500 nm.

**4. Polariton-enhanced molecular spectroscopy**

We investigated polariton-enhanced spectroscopy for nanomaterials and biomolecules using the reported Si metasurface. To demonstrate its versatility, we first performed DUV metasurface-enhanced double resonance Raman scattering measurements to for monolayer tungsten disulfide (2D-WS$_2$) and molybdenum disulfide (2D-MoS$_2$). They were selected as analytes due to their relevance as key materials in next-generation electronic[62-63] and optoelectronic devices[64-65]. Their atomically thin nature makes them ideal candidates for testing the sensitivity of our metasurface to ultrathin nanomaterials.

Resonance Raman scattering[66-67] occurs when the photon energy matches an electronic transition of the analyte, significantly enhancing Raman signals. Additionally, double resonance Raman scattering arises when intervalley transitions, coupled with electron-phonon interactions, contribute to the scattering process, resulting in additional spectral features. This phenomenon provides valuable insights into the electronic band structure, phonon dispersion, and electron-phonon coupling in nanomaterials, making it highly beneficial for material physics investigations. The monolayer analytes were transferred onto two separated metasurfaces. Atomic force microscopy (AFM) images confirmed the successful transfer, as shown in **Figures S9**a and S9b in Supporting Information. For Raman measurements in the DUV and visible, continuous wave (cw) lasers with operation wavelengths of 266 nm and 488 nm, respectively, were used. The measurement results were summarized in **Figure 3**. In both cases, compared to the Raman spectra of the monolayers on the adjacent bare Si substrates (Figures 3a and 3c), we observed significant enhancement of Raman signals on the metasurface areas under DUV excitation, confirming the enhancement capability of the reported metasurface. As shown in Figure 3d, the signals of MoS$_2$, which are much weaker compared with the Si substrate's Raman signal at 520 cm$^{-1}$, can be overlooked under the visible excitation, even with extended integration times. In contrast, in the metasurface-enhanced spectra, the signals of the 2D materials can be efficiently observed.

In detail, for 2D-WS$_2$, Raman peaks at 356 cm$^{-1}$ and 422 cm$^{-1}$, corresponding to the $E_{2g}^1(\Gamma)$ and $A_{1g}(\Gamma)$ modes, were enhanced by more than 14-fold and 7.5-fold, respectively, compared to the bare Si area under DUV excitation. Notably, the metasurface's enhancement performance in the DUV regime is similar to state-of-the-art plasmonic devices[1, 22, 68] previously used for biomolecular analysis. Compared with the visible Raman spectrum (Figure 3b), additional peaks at around 707, 763, and 825 cm$^{-1}$, attributed to double resonance Raman scattering, were also observed in the metasurface-enhanced DUV Raman spectrum. These peaks, according to Kawata *et al.*[69], associated with modes $E_{2g}^1(M) + E_{2g}^2(M)$, $A_{1g}(M) + E_{2g}^2(M)$, and $2A_{1g}(M)$, respectively, provide detailed insights into the electronic band structure and phonon dispersion of the material. Notably, these features were barely visible or entirely absent on the bare Si surface, further highlighting the metasurface's enhancement capability. For 2D-MoS$_2$, DUV excitation revealed sharp Raman peaks at ~383 and 407 cm$^{-1}$, corresponding to its characteristic vibration modes of $E_{2g}^1(\Gamma)$ and $A_{1g}(\Gamma)$. Additional weak signals at around 725, 758, 788, and 819 cm$^{-1}$, which are not observable in the visible Raman spectrum (Figure 3d) and correspond to modes[70] of $A_{1g}(M) + E_{1g}(M)$, $2E_{2g}^1(M)$, $2E_{2g}^1(\Gamma)$, and $2A_{1g}(\Gamma)$, were observed respectively.

In addition, the Si metasurface exhibits negligible photoluminescence (**Figure S15a**, Supporting Information), providing a low-noise background—a critical advantage compared to substrate materials such as silica, which often present high background noise in DUV Raman measurements. Furthermore, we observed distinct modulation of the Si Raman signal at 520 cm$^{-1}$ when capped with different 2D materials: the Si's signal was enhanced with MoS$_2$ but not with WS$_2$. This discrepancy likely arises from variations in optical absorption or coupling strength between the metasurface and different 2D materials in the DUV regime. The underlying mechanism remains unclear, presenting an intriguing opportunity for future investigation to better understand light-matter interactions in such hybrid systems.

With its ability to enhance resonance Raman scattering for nanomaterials and its low noise background, we propose the reported Si metasurfaces as an ideal platform for DUV Raman spectroscopy. Overall, the results highlight the metasurface's potential to significantly enhance resonance Raman scattering in monolayer 2D materials, achieving sensitivity to atomic-scale thickness. These findings confirm the utility of the reported DUV metasurface for material differentiation and analysis, with promising applications in clinical diagnostics and biomedical research in the future[1, 15].

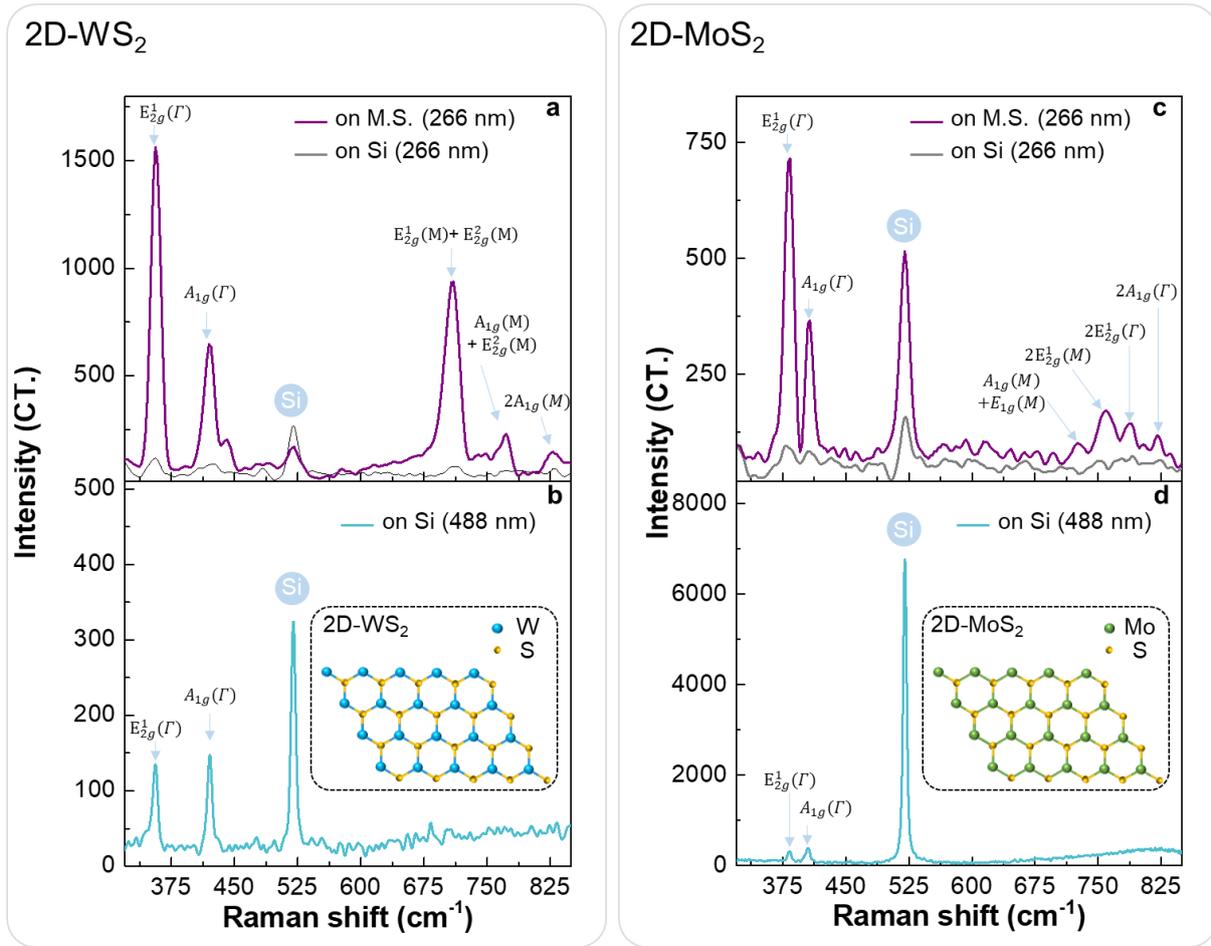

**Figure 3.** Polaritonic-metasurface-enhanced double resonance Raman spectroscopy. a) DUV Raman spectra acquired from the transferred 2D-WS$_2$ on the metasurface (denoted as "on M.S.") and adjacent bare Si surface (denoted as "on Si"). b) 488-nm Raman spectrum of the transferred 2D-WS$_2$. Inset shows the structure of the 2D-WS$_2$. CT.: counts of the collected photons. c) DUV Raman spectra acquired from the transferred 2D-MoS$_2$ on the metasurface (denoted as "on M.S.") and adjacent bare Si surface (denoted as "on Si"). d) 488-nm Raman of the transferred 2D-WS$_2$. Inset shows the structure of the 2D-WS$_2$. CT.: counts of collected photons.

Next, we explored the metasurface's capability for autofluorescent biomolecular spectroscopy (**Figure 4a**) in solution, which is critical for biomedical research and sensing applications, as biomolecules often require their native environment for accurate analysis. Autofluorescence is the intrinsic property of many biomolecules, most of which emit fluorescent light by DUV excitation. Autofluorescence has numerous applications in biomolecular analysis and disease research, as it is useful for molecule identification and tracking[71]. For example, nicotinamide adenine dinucleotide (NADH), where the molecular structure is shown in Figure 4b, is a critical coenzyme majorly associated with cellular metabolism[72], while its autofluorescence has been found to be a vital indicator for tumor development and aging disease[73-74]. Here we demonstrate that our Si metasurface efficiently enhances NADH autofluorescence. For the spectral tests, we dissolved NADH in PBS with a concentration of 0.015 mM. Figure 4c shows the UV-visible absorbance spectrum of the solvent. Two absorbance peaks centered at ~260 nm and ~337 nm, respectively, are identified in the spectrum. The absorbance peak at ~260 nm corresponds to the adenine moiety of the NADH molecule[75]. Due to the proper overlap between the biomolecule's first absorption band and the metasurface's resonance wavelength, the resonant near field generated by the metasurface at 266 nm can effectively enhance the excitation[76] of NADH, resulting in stronger autofluorescence emission. The native oxide layer formed on Si in the metasurface serves as a thin and uniform dielectric spacer that prevents direct contact between the metasurface and the excited molecules. This oxide layer and the dielectric properties of Si in the molecular emission band prevent the quenching effects due to contact with metals[77-79]. We immersed the metasurface sample into the solvent and used a confocal spectroscope equipped with a 266 nm continuous-wave (cw) semiconductor laser as the excitation for the measurement. The confocal scheme of the spectrometer ensures the minimum influence of the out-of-focus fluorescence signal from the background. We also simulated the field enhancement of the polaritonic metasurface in the solution and confirmed that it still provides a proper field enhancement at 266 nm (Figure S10 in the Supporting Information). To make a comparison, we measured the autofluorescence spectra from the NADH solvent on the metasurface (denoted as "on M.S.") and adjacent bare Si surface (denoted as "on Si"), where the results are presented in Figure 4d. The results clearly show that the metasurface enhances autofluorescence over the emission band from 350 to 600 nm. This can also be confirmed by calculating the enhancement factor, defined as the intensity ratio between the NADH signal from the metasurface area ($I_{M.S.}$) and the signal from the bare Si area ($I_{Si}$), where the result is plotted in Figure 4e. We calculated the integrated autofluorescence intensity collected from the metasurface and bare Si areas and made a comparison. The NADH on the metasurface shows a ~3.2-fold enhancement of the overall intensity compared with the NADH molecules on the bare Si area. The enhancement factors provided by the reported polaritonic Si metasurfaces are comparable to the previously reported plasmonic devices[17, 80-81]. **Figure S13** in Supporting Information shows a photographic image of the sample, which indicates that with the enhancement of the metasurface, the autofluorescence from NADH can be clearly observed (the blue spot pointed by the arrow) by a general camera. In addition, after the measurement, the metasurface was cleaned and observed using SEM. No apparent laser-induced damage or photon corrosion was detected, indicating that the Si metasurface provides a robust platform for future DUV spectroscopy measurements in solvents **(Figure S14).** We performed photoluminescence measurements on the Si chip with the 266 nm excitation and did not detect any significant photoluminescence signal **(Figure S15**a**)**. To compare, we also measure the photoluminescence of a 300-nm-thick AlN thin film. Despite its low optical absorption in the DUV and suitability for fabrication of Mie-type metasurfaces, it shows a very bright photoluminescence (Figure S15b), with signal much stronger than NADH solution on the Si substrate. Thus, it may not be suitable for DUV sensing metasurface application as the intrinsic photoluminescence of AlN will cause strong background noise and obscure the spectral signals from the analyte. The result confirms the ability of Si to provide a spectrally clean background for sensitive spectral measurements, which is essential in spectroscopy techniques such as Raman and photoluminescence measurements of biomolecules and nanomaterials. These results confirm the metasurface's capability to enhance biomolecular autofluorescence and highlight its potential for relevant biological research and clinical diagnosis.

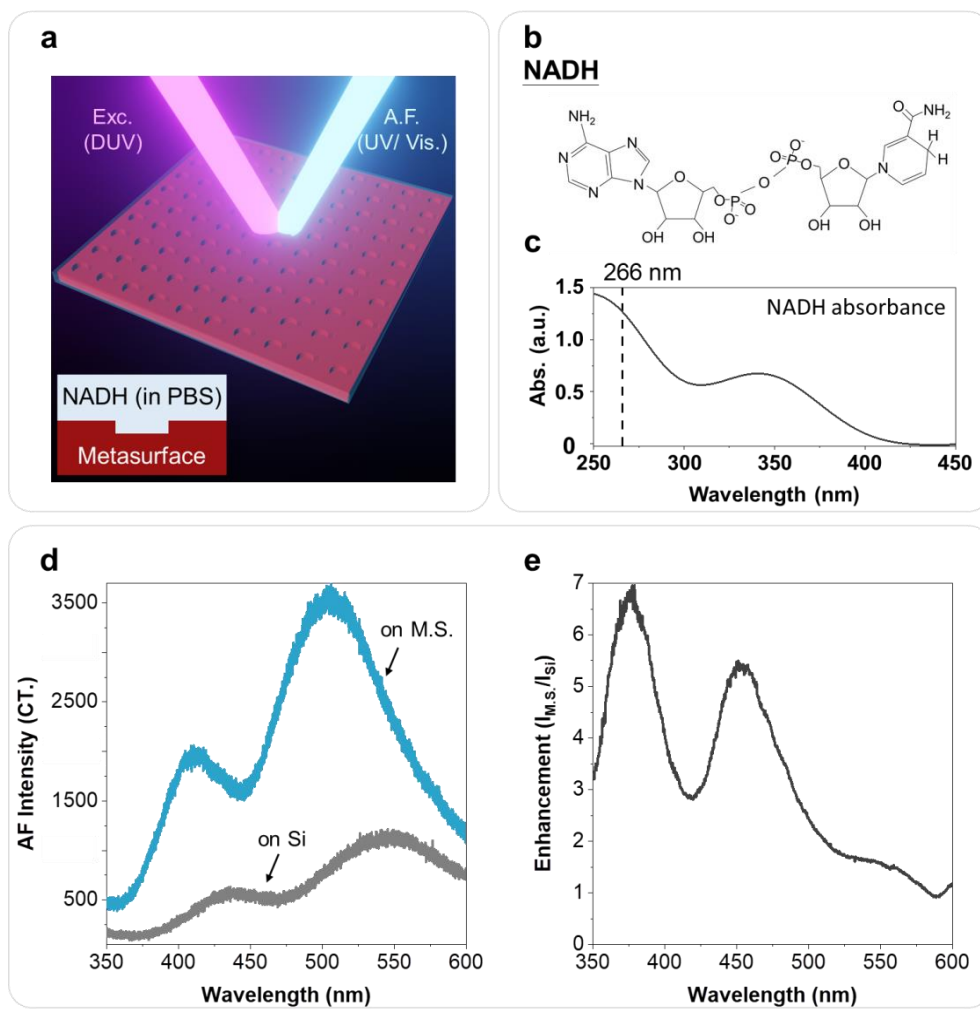

**Figure 4**. Polariton-enhanced autofluorescence of biomolecule NADH. a) Schematic of the experiment. Exc.: Excitation, A.F.: autofluorescence. b) Molecular structure of NADH. c) Absorbance spectrum of NADH solution. d) Autofluorescence spectra of the NADH observed on the metasurface (M.S., green line) and the adjacent bare Si surface (Bare Si, gray line). CT.: counts of the collected photons. e) Enhancement factor by dividing the autofluorescence signal on the metasurface area by the autofluorescence signal from the bare Si area.

## 5. Stability test

Finally, we test the stability of the Si metasurface in solvents commonly used in biological protocols. In many clinical and sensing processes, it is necessary to immerse the sensing chips in different solvents for analyte binding or surface functionalization[7]. Testing the stability of the reported metasurfaces in the solvents can further verify their potential for relevant applications. For comparison, we produced Al metasurfaces (**Figure S17** in Supporting Information) using focused ion beam milling, which shows a resonance at approximately 280 nm. The period of the Si metasurface was slightly changed to 220 nm, resulting in the resonance wavelength moving to ~245 nm. This reduction in the period also decreased the minimum gap size between voids, making the metasurface more sensitive to potential damage and allowing for clearer observation of any structural changes from monitoring the associated spectral variation.

We tested the stability of the samples in three solvents: phosphate-buffered saline (PBS, pH 7.4), Tris-EDTA buffer solution (pH 7.6), and phosphoric acid ($H_3PO_4$, pH 6.5). We prepared three samples and immersed them in the solvents for the stability test for both Si and Al metasurfaces. After specific time periods, we removed the samples from the solvent, rinsed them with deionized (DI) water, dried them with nitrogen gas purging, and performed spectral measurements. Overall, in the spectral measurements (red curves in **Figure 5**), the Si metasurfaces demonstrated better robustness to the solvents: The resonance dip for each sample doesn't deteriorate significantly after the immersion in PBS (red curve in Figure 5a), Tris-EDTA (red curve in Figure 5b), and $H_3PO_4$ (red curve in Figure 5c). The Si metasurfaces show a resonant dip at ~250 nm. The slight spectral differences between the samples can be associated with the imperfection in each sample. After 24-hour immersion in Tris-EDTA buffer, the resonance dip becomes slightly broader and red-shifted. It is found that a thin film was deposited on the top of the sample. Although the film formation mechanism is unclear yet, it was removable using a standard RCA cleaning process[82]. After the cleaning, the resonant dip (labeled as cleaned in Figure 5b) returns to the original spectral position before 24-hour sample immersion. In contrast, the Al metasurfaces were damaged by PBS, $H_3PO_4$, and Tris-EDTA buffer solutions within 24 hours, 1.5 hours, and 1.5 hours, respectively, as can be identified from the spectral measurements (the gray curves in Figure 5). These results indicate that the reported Si metasurface is a stable platform for DUV molecular spectroscopy measurements that require repetitive immersing the sensing chips in solvents[83].

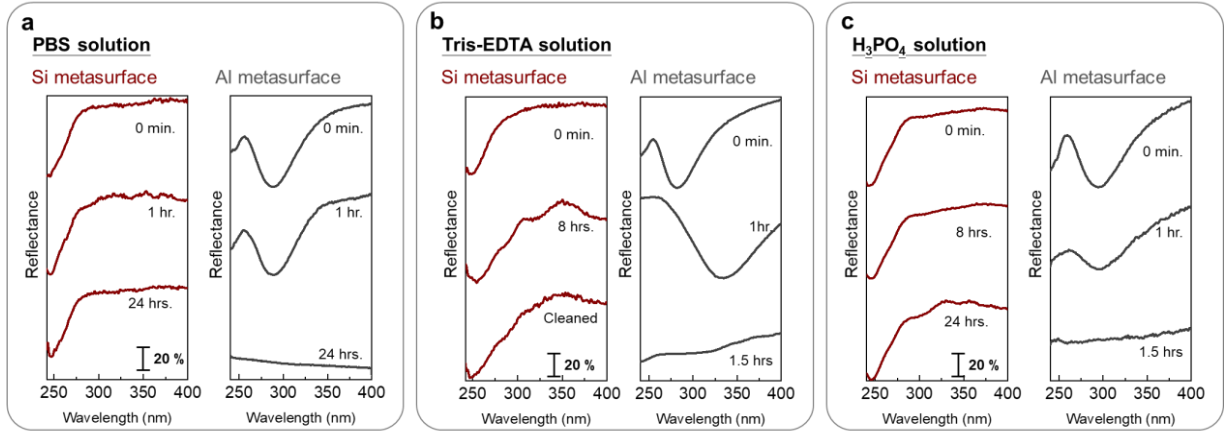

**Figure 5.** Stability test and comparison. The reflectance spectrum of Si (red curves) and Al (gray curves) metasurfaces after being immersed in a) PBS, b) Tris-EDTA and, c) $H_3PO_4$ solutions for different time periods.

## 6. Conclusion

We developed a DUV Si metasurface by harnessing interband-transition-induced polaritonic resonance and fine-tuning the balance of multipole resonances in void-type unit cells. With deep subwavelength field confinement, low intrinsic photoluminescence background noise, and exceptional stability, the Si polaritonic metasurface enhances the interactions between DUV light and analytes and boosts the spectroscopy signals, showcasing its potential for advanced DUV nanophotonic sensing.

For biological applications, our metasurface achieves in-solution biomolecule autofluorescence enhancement. The result opens possibilities for future label-free autofluorescence imaging, molecular tracing, and disease diagnosis by using the reported metasurface. To further expand the Si polaritonic metasurface's enhancement capabilities, designing unit cells with multiple resonance modes—one exhibiting polaritonic resonance in the DUV and another featuring a low-loss Mie mode—could enhance biomolecule autofluorescence from its excitation and emission[84-85] bands, respectively[76]. Integration with microfluidic systems[86] and surface functionalization techniques[7] could enable real-time monitoring of biomolecular interactions in solution, paving the way for compact and efficient biosensing devices that combine sensitivity, selectivity, and stability.

Our metasurface enhances double-resonance Raman scattering in monolayer transition metal dichalcogenides, demonstrating its potential for probing the electronic structures and phonon dispersions of nanomaterials. Future works combining the metasurface's capabilities with advanced theoretical approaches, such as density functional theory (DFT)[69], could provide deeper insights into the electronic and crystal structures of emerging materials.

Enabling applications in monitoring material synthesis, phase transitions, and other dynamic processes can also be envisioned.

Additionally, recent results[38] reveal that Si maintains polaritonic properties up to the vacuum UV (VUV, ~11 eV) band, showing its promising potential for VUV polaritonics. Given the extensive use of VUV in material research[87-88] and biomedical analysis[89], this study lays a foundation for advancing polariton-enhanced UV spectroscopy. Beyond these applications, Si's abundance, scalability, and compatibility with mature CMOS fabrication techniques position it as a cost-effective platform for practical DUV sensing devices. Scaling up fabrication using CMOS-compatible processes[90-91] will facilitate widespread adoption, while more sophisticated resonance mechanisms in unit cell designs can further enhance performance.

In conclusion, our work establishes a robust platform for the next generation of DUV spectroscopy technologies. By leveraging Si's unique properties, scalability, and interdisciplinary applicability, this study bridges gaps across nanotechnology, biology, and material science, paving the way for key research and practical innovations.

## 7. Experimental Section

**Design**

We used FDTD-based simulation software Lumerical to design and analyze the properties of the metasurfaces presented in this work. In the simulation, periodic boundary conditions were imposed along the *x*- and *y*-axis, and PML boundary conditions were used along the *z*-direction. The incident light is linearly polarized. The minimum mesh size is 1 nm in the critical regions in the simulation model. The optical constants of Si were adopted from Ref. [47]. In the calculation of the multipole decomposition, we calculated the electric displacement induced in the unit cell and used it for calculating the strength of the individual multipoles. For the implementation, we used the MATLAB code reported in Ref. [53]. More details can be found in Supporting Information.

**Sample fabrication**

We used electron beam lithography to make the Si metasurfaces (see **Figure S18** in the Supporting Information for the process flow). A 90-nm-thick polymethyl methacrylate (PMMA 950 A2) film was spin-coated on a cleaned Si chip and baked at 180 °C for 2 minutes. In this fabrication, the PMMA layer serves a dual purpose: it defines the metasurface pattern during the e-beam exposure and acts as the etching mask for the etching process. This approach eliminates the need for additional metal layer evaporation and the lift-off process, thereby enhancing fabrication efficiency. The exposure was performed by using a pattern generator (Raith VOYAGER, acceleration voltage: 50 kV, current 0.4 nA). The exposed sample was immediately immersed into MIBK/IPA (ratio 1:3) and rinsed in IPA for 75 and 30 seconds, respectively, for the development process. Subsequently, dry etching was performed by using an inductively coupled plasma etching system (ICP-RIE, LAM research, Model 2300) to transfer the pattern into the Si chip. The gas mixture for Si etching consists of $Cl_2$ and HBr. The PMMA layer was removed by using acetone. To correct the distortion of the resulting patterns caused by the proximity effect, the layout file used for the e-beam exposure was modified, see **Figure S8** in the Supporting Information. To make the Al metasurfaces, 40-nm-thick Al films were evaporated on Si chips. The base pressure is $\sim 5 \times 10^{-6}$ Torr, and the evaporation rate is ~1 nm/s. The Al films were nanopatterned by using a focus ion beam system (Helios 660 NanoLab, FEI). The acceleration voltage and beam current of the gallium ion beam were 30 kV and 50 pA, respectively, while the dose for the process was 210 $\mu C/cm^2$. A commercial software NPGS (JC Nabity Lithography Systems) capable of precisely controlling the ion beam path and the dose was used in the process. The sample areas of the Si and Al metasurfaces are 80×80 and 50×50 $\mu m^2$, respectively.

**Spectral measurement**

To perform the optical characterization, we built a DUV-visible micro-spectrometer (see **Figure S19** in the Supporting Information for the system's schematic). A broadband laser-driven plasma light source (ISTEQ, XWS-30, wavelength range: 190nm-2500 nm) was used as the illumination for the reflectance measurement. Three Al-coated off-axis parabolic mirrors were used to shrink the spot size and make the light collimated. A polarizer cube (Thorlabs RPM10) is used to control the polarization of the incident light. Multiple UV-grade Al mirrors were used for the alignment (from Thorlabs). The sample was mounted on a three-dimensional linear stage to control the precise position of the sample. For the Si measurement, a clean Si chip was used as the reference to measure the relative reflectance of the metasurface sample, similar to the approaches reported in Ref. [92]. Multiple spectra were collected for each measurement and averaged for the data presented in the paper. Similarly, for the Al metasurface measurement, a clean Al mirror was used as the reference. An UV achromatic objective lens (50× Mitutoyo plan UV infinity corrected objective) was used for the imaging, excitation, and signal collection. A spectrometer (OtO Photonics, EagleEye EE2063, wavelength resolution: 0.2 nm) was used for the spectral analysis. To restrict the range of the incident angle, an additional pinhole with a diameter of 2 mm was mounted on the front end of the objective lens. The spot size of the incident light is around 40 μm in diameter.

**TMD sample growth and transfer**

$WS_2$ and $MoS_2$ flakes were synthesized by chemical vapor deposition that uses transition metal oxide powder and sulfur powder as the precursors. They grow on sapphire substrates at 900 °C. After growth, a PMMA membrane was spin-coated on the 2D material flakes, followed by thermal treatment at 100 °C for crosslinking of PMMA with the film. Next, polydimethylsiloxane (PDMS) and deionized water were used as transferring medium, and full delamination (peeling off) of the 2D material flakes from the sapphire substrates. Because the sapphire substrate is hydrophilic and the TMD films are hydrophobic, this action can separate the 2D material from the sapphire surface, and the PMMA/2D material bilayer is released and floating on water. Next, the PMMA/2D material on water is fished up onto the surface of the Si sample and then air-dried. To avoid delamination of the 2D materials from the patterned Si substrate during the PMMA removal, we anneal the sample in Argon flow at 400 ℃ for 4 hours to burn off PMMA.

**Autofluorescence and Raman measurement**

A micro Raman spectrometer (Horiba Jobin-Yvon LabRam HR800) was used for the autofluorescence and resonance Raman measurement. A 266-nm continuous-wave semiconductor laser (FQCW 266-50, Crystal Laser Systems) was used as the excitation. The system emits a linearly polarized DUV laser beam. A 40× DUV objective (LMU-40X-UVB, working distance: 0.8 mm, numerical aperture (NA): 0.49) was used for the excitation and signal collection. For the DUV Raman measurement, a grating with a groove density of 1800 gr/mm was used. The laser's polarization was carefully aligned with the sample during spectroscopy measurements, oriented along the neck of the fish-shaped void unit cell. The incident power is 2.25 mW.  For the visible Raman measurement, a 488-nm laser (Stellar-Pro, Modu-Laser, LLC) was coupled into the spectrometer system and used as the excitation.

**Supporting Information**

Supporting Information is available from the Wiley Online Library or from the author.

**Acknowledgements**

We would like to thank Prof. Shi-Wei Chu for his helpful comments and suggestions on this paper. Funding: The authors acknowledge support from the National Science and Technology Council (NSTC 112-2636-M-A49-001, NSTC 113-2636-M-A49-001) in Taiwan, the Ministry of Education in Taiwan under the Yushan Fellow Program. This work was also supported by the Higher Education Sprout Project of National Yang Ming Chiao Tung University and the Ministry of Education (MOE), Taiwan. The authors are grateful to the support from the Center for Integrated Electronics-Optics Technologies and Systems in National Yang Ming Chiao Tung University, and the Taiwan Semiconductor Research Institute (TSRI). We are grateful to Research Center for Applied Sciences, Academia Sinica, for their support in using the FIB system (FEI Helios 660 Nanolab) and the SEM system (FEI Inspect F SEM). YK acknowledges support from the Australian Research Council (Grant No. DP210101292) and the International Technology Center Indo-Pacific (ITC IPAC) via Army Research Office (contract FA520923C0023). Author contributions: MLT conceived the idea and designed the research; BRL performed the simulations, designed the samples, and optimized the fabrication process and the measurement. DHL assisted the setup of the spectrometer and paper preparation. MHS assisted the focus ion beam milling of the metasurface. YCP, and MFC fabricated the Si sample. SCC performed AFM measurement. PYH, KHC, and JHL performed the spectral measurements. BRL, MFC, PHH, and CHC assisted the sample preparation for the Raman and autofluorescence measurement. MFC and PHH performed the Raman and autofluorescence measurement. TEL, YCL, QYL, CTL, RHH helped the experiment designs for Raman measurements. MLT, BRL, and YK wrote the paper. YK and MLT supervised the project.

**Conflict of Interest**

The authors declare no conflict of interest.

**Data Availability Statement**

The data that support the findings of this study are available from the corresponding author upon reasonable request.

**50-word table of contents entry**

A stable silicon metasurface leveraging polaritonic resonances and Kerker-type modes enhances DUV molecular spectroscopy, improving sensitivity for biomolecule autofluorescence and double-resonance Raman scattering in 2D materials while overcoming the limitations of conventional plasmonic structures.

B.-R. Lee, M. F. Chiang, P. Y. Ho, K.-H. Chen, J.-H. Lee, P. H. Hsu, Y. C. Peng, J.-Y. Hou, S.-C. Chen, Q.-Y. Lee, C.-H. Chang, B.-R. Li, T.-E. Lin, C.-T. Lin, M.-H. Shih, D.-H. Lien, Y.-C. Lin, R.-H. Horng, Y. Kivshar, M. L. Tseng*

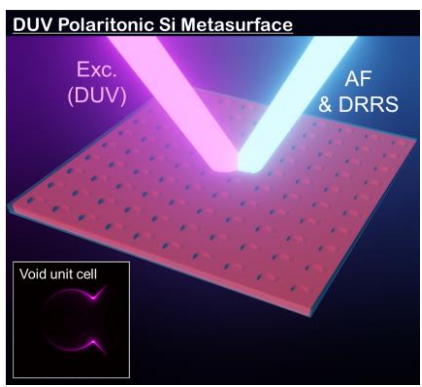